\documentclass[usegraphicx,useAMS,usenatbib]{mn2e}
\include{latex_macros}

\title[Random alignment of spin axes?]
 {Are the spin axes of stars randomly aligned within a cluster?}

\author[R. J. Jackson and R. D. Jeffries]
  {R. J.~Jackson and R. D.~Jeffries\\
  Astrophysics Group, Research Institute for the Environment, Physical
  Sciences and Applied Mathematics, Keele University, \\ Keele, 
      Staffordshire ST5 5BG
}
\date{Submitted August 12 2009}

\pagerange{\pageref{firstpage}--\pageref{lastpage}} \pubyear{2009}

\def\LaTeX{L\kern-.36em\raise.3ex\hbox{a}\kern-.15em
    T\kern-.1667em\lower.7ex\hbox{E}\kern-.125emX}

\begin{document}
\label{firstpage}
\maketitle

\begin{abstract}
We investigate to what extent the spin axes of stars in young open
clusters are aligned. Assuming that the spin vectors lie uniformly
within a conical section, with an opening half-angle between
$\lambda=0^{\circ}$ (perfectly aligned) and $\lambda=90^{\circ}$
(completely random), we describe a Monte-Carlo modelling technique that
returns a probability density for this opening angle given a set of
measured $\sin i$ values, where $i$ is the unknown inclination angle
between a stellar spin vector and the line of sight. Using simulations
we demonstrate that although azimuthal information is lost, it is
easily possible to discriminate between strongly aligned spin axes and
a random distribution, providing that the mean spin-axis inclination
lies outside the range $45^{\circ}$--$75 ^{\circ}$.  We apply the
technique to G- and K-type stars in the young Pleiades and Alpha Per
clusters. The $\sin i$ values are derived using rotation periods and
projected equatorial velocities, combined with radii estimated from the
cluster distances and a surface brightness/colour relationship. For
both clusters we find no evidence for spin-axis alignment:
$\lambda=90^{\circ}$ is the most probable model and $\lambda >
40^{\circ}$ with 90 per cent confidence. Assuming a random spin-axis
alignment, we re-determine the distances to both clusters, obtaining
$133\pm 7$\,pc for the Pleiades and $182\pm 11$\,pc for Alpha Per. If the assumption of random
spin-axis alignment is discarded however, whilst the distance estimate remains
unchanged, it has an additional $^{+18}_{-32}$ percent uncertainty.
\end{abstract}


\section{Introduction}
Most authors considering the statistics of orbital or rotational
stellar motion assume, where relevant, that angular momentum vectors
are randomly orientated.  It is possible however that the physical
processes of star formation lead to a preferred axis of rotation over
the scale of an individual star forming region (SFR). This might arise
if the direction of average angular momentum of the molecular cloud
giving rise to the SFR has a significant influence on the resulting
angular momentum of individual stars -- for instance, if gas were
constrained to collapse along strong, large-scale magnetic fields
threading the cloud.

Historically, little has been discussed either theoretically or
observationally about the possibility of spin-axis alignment during
star formation. This would require a relatively undisturbed collapse
along magnetic field lines with little disruption from turbulence or
dynamical interactions (e.g. Shu, Adams \& Lizano 1987).  Using
circumstellar disc orientation as a proxy, some studies have suggested
preferential spin alignment with the ambient magnetic field in SFRs
(e.g. Tamura \& Sato 1989; Vink et al. 2005), but others have found no
evidence for disc axis alignment (M\'enard \& Duch\^ene 2004).

A second important reason for assessing the degree of spin-axis
alignment is that measurements of projected equatorial rotation
velocities ($v \sin i$, where $i$ is the unknown inclination of the
spin-axis to the line of sight) and rotation periods can be combined to
provide a powerful method to determine the distances (Hendry, O'Dell
and Collier-Cameron 1993; Jeffries 2007a; Baxter et al. 2009), radii (Jackson, Jeffries \&
Maxted 2009) or star formation histories (Jeffries 2007b) of young
clusters. Such statistical analyses {\it must assume} that the
orientation of spin axes are random.  If the spin axes in an individual
cluster were in fact partially aligned, this would change the intrinsic
$\sin i$ distribution for the cluster, producing biased estimates of
distances, radii and age spreads.

In this paper we use published rotation data for the young Pleiades and
Alpha Per open clusters to investigate to what extent spin axes may
be aligned once the star formation process has finished.  In Section 2
we discuss how well spin-axis orientation can be determined using
measured rotation periods and projected equatorial velocities. In
Section 3 we present a parameterised model for the observed $\sin i$
distribution resulting from a group of stars with partially aligned
spin axes and show how well such a model can be used to determine the
degree of alignment from simulated datasets. In Section 4 we compare the
models with measured $\sin i$ distributions for G- and K-type stars in
the Pleiades and Alpha Per clusters.  Section 5 presents and discusses
the results of our analyses, including new, independent estimates of
the distances to these clusters under the assumption that the spin axes
{\it are} randomly aligned, and in Section 6 we give our conclusions.

\section{Observation of spin axes}
Using current measurement techniques it is not possible to observe
stellar spin-axis orientation directly. For young, magnetically active, spotted
late-type stars it is possible to measure their period
of rotation, $P$, from rotational modulation of their light curves 
and their projected equatorial velocity, $v\sin
i$, from spectral line broadening. 
These can be used together with photometric data and an
independent measure of distance to determine $\sin i$, the sine of the
angle between the observer and the spin axis
\begin{equation}
\sin i = \frac{P}{2\pi R}\, (v\sin i)\, , 
\end{equation}
where the stellar radius $R$ can be estimated from the surface
brightness and distance. Such observations give no
information on the azimuthal direction of the spin axis, only an
estimate of the
inclination, $i$, which is also degenerate 
between $i$ and $\pi - i $.

The effect of restricting observations to measurements of $\sin i$ is
illustrated in Fig.~1. The left-hand panels indicate the intrinsic
distribution of spin-axis vectors. The central panels show what would
be observed given the lack of azimuthal information and the
degeneracy in $i$ discussed above. The right hand panels show the
cumulative $\sin i$ distributions that would be observed (see section~3.1).

In the first case, where there is no preferred orientation, the uniform
distribution of spin axes would be observed as a uniform inclination
distribution
over a hemisphere from 0 to $\pi/2$. The next case is more
complicated. Here, the spin axes of a group of stars are distributed
uniformly over a conical region about a central cone axis with defined
inclination and azimuthal direction. This would be observed as a larger
circular region on the surface of the hemisphere that is symmetric
about the line of sight; the azimuthal information being lost.

The third and fourth cases show how conical regions representing the
same degree of alignment would be observed at increasing average
inclinations.  The lack of azimuthal information causes spin axes
represented by a relatively small cone area to sweep out a large area
of the measured hemisphere. This concept of swept areas is
approximate. A detailed treatment is given below, where the variation
in number density over the swept area is modelled to calculate the
cumulative probability density as a function of the opening angle of
the cone, $\lambda$, and its mean inclination, $\alpha$. The point here
is to emphasise that azimuthal and inclination angle degeneracies
conspire, especially when convolved with measurement errors (see next
section), to hamper the recovery of the underlying $\sin i$
distribution.

\begin{figure*}
\centering
\begin{minipage}[htb]{0.9\textwidth}
\includegraphics [width = 150mm]{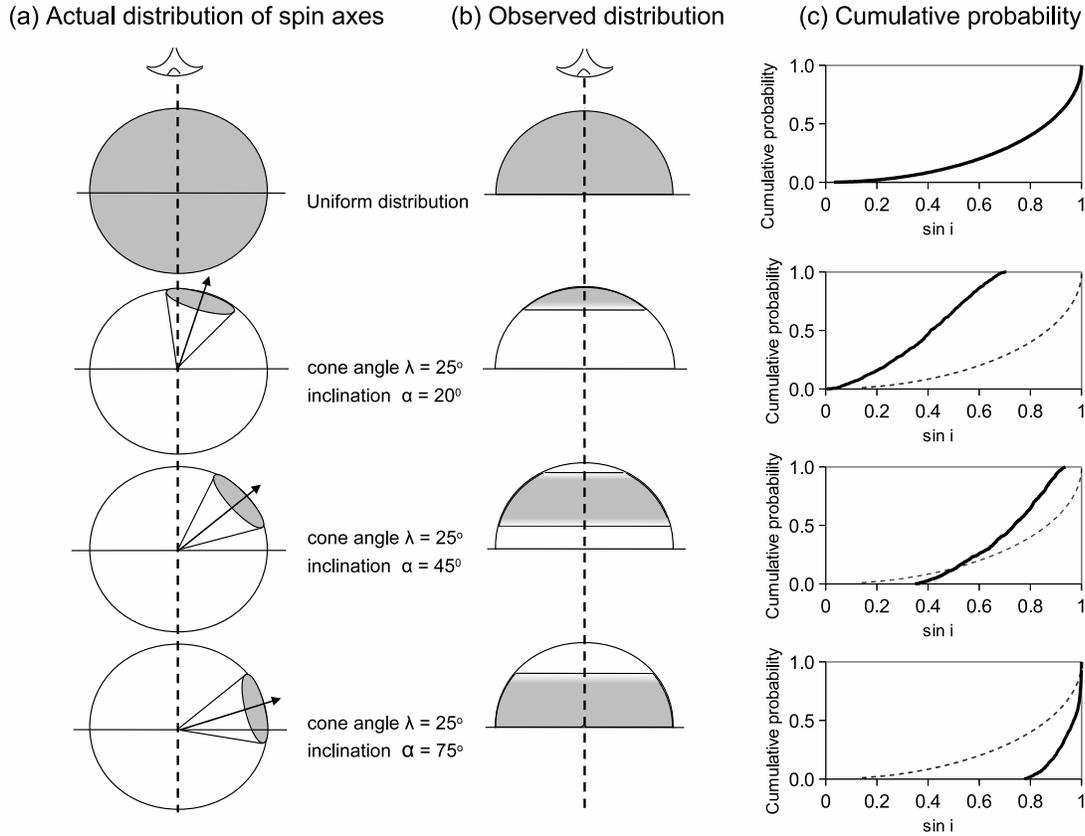}
\end{minipage}
\caption{A schematic diagram showing (a) the distribution of spin
  vector orientations, (b) the corresponding
distribution that would be deduced from $\sin i$ measurements and (c)
the resultant cumulative probability distribution of measured $\sin i$ (solid line)
compared to that found for a uniform distribution of spin axes (dashed line).
}
\label{schematik}
\end{figure*}

\section{Modelling the \mbox{${{\large \mathbf \sin i}}$} distribution}
We will use the term ``$\sin i$ distribution'' for a group of stars to
express the set of numbers specifying the angle between the rotation axis 
and the observer's line of sight. Two such distributions can be
considered:
\begin{enumerate}
\item \indent a true $\sin i$ distribution which depends only on the
distribution of the spin axes of a set of stars,
\item \indent a measured $\sin i$ distribution which depends both on
the true distribution and the uncertainties and limits that apply to 
measurements of the base parameters used to determine $\sin i$.
\end{enumerate}

\subsection{The true $\sin i$ distribution}
The true $\sin i$ distribution depends on the distribution of the stellar
spin axes over the celestial sphere. The simplest case is 
a uniform (random) distribution in which case the cumulative probability
distribution depends on the area of the celestial sphere between an
angle 0 and $i$
\begin{equation}
P^{o}_{\rm true} = 1 - \cos i\;\;\;    {\rm for}\; i = 0 \; {\rm to}\;
\pi/2\, ,	
\end{equation}
where the superscript $o$ denotes the case of a uniform distribution. 

A simple way to 
represent an aligned distribution is to assume that spin vectors are uniformly
distributed over a conical solid angle and zero elsewhere (see
Fig.~1). The cone angle, $\lambda$, which corresponds to half the
opening angle, determines the degree of alignment. A small cone angle
means stars have nearly parallel spin. A large cone angle ($\lambda
\approx \pi/2$) corresponds to a uniform distribution. The mean
inclination of the stars within the cone is represented by $\alpha$.

\begin{figure*}
\centering
\begin{minipage}[tb]{0.9\textwidth}
\includegraphics [width = 150mm]{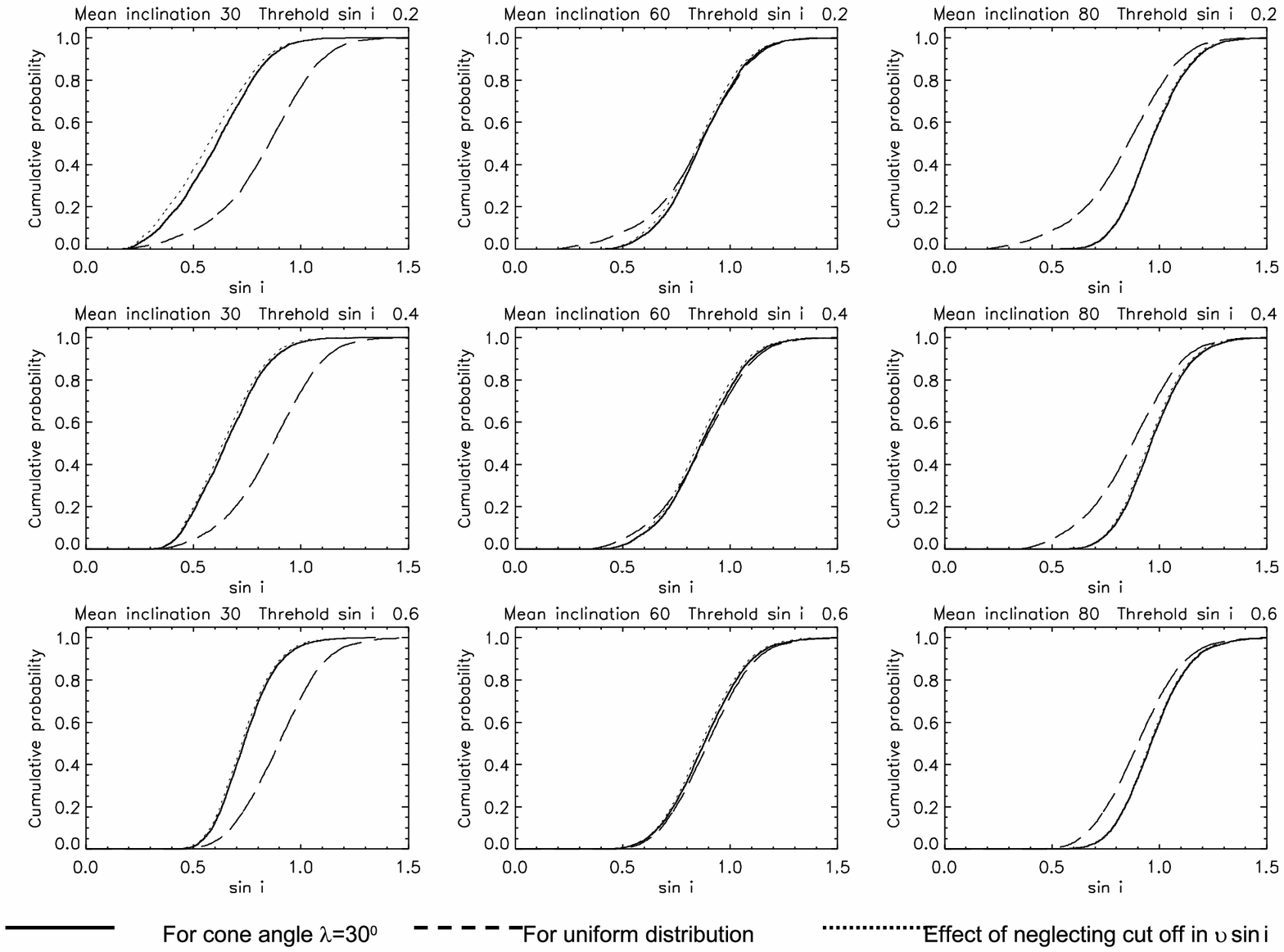}
\end{minipage}
\caption{Simulations of the cumulative probability distributions of observed
$\sin i$ assuming normalised uncertainties $\delta_{PV}=\delta_{AD}=0.1$ and
a distribution of spin-axis orientations that is either a
uniform (random) or is aligned
within a cone angle  $\lambda=30^{\circ}$.  The  rows show results for 
mean  inclinations of $\alpha=30^{\circ}$,
$60^{\circ}$ or $80^{\circ}$. The columns show the effects of
altering $\sin i_{\rm min}$. The dotted line shows the effect of
setting $(v\sin i)_{\rm min}=0$. Note the value of $\sin i$ can exceed 
unity due to uncertainties in the measured parameters used to determine $\sin i$.}
\label{cdf}
\end{figure*}
 
The equivalent $\sin i$ distribution can then be calculated using a Monte
Carlo method as follows. A set of spin axes are specified at angles 
$\theta_{n}$ and $\phi_{n}$,
{\it relative to the cone axis}. The cumulative probability
distribution depends on the angle between the cone axis and spin axis,
$\theta_{n}$, as
\begin{equation}
P (\theta) = (1 - \cos \theta)/(1-\cos \lambda)\;\;\; {\rm for}\, \theta
= 0\; {\rm to}\;\lambda			
\end{equation}
and random values of $\theta_{n}$  are generated as 
\begin{equation}
\theta_{n} = \cos^{-1} (1 - R_{n} (1-\cos\lambda)) 
\end{equation}
where $R_{n}$ is a random number between 0 and 1. The probability
distribution of the angle $\phi$ around the cone axis is uniform,
allowing random values of $\phi$ to be generated as $\phi_{n} = 2{\pi}
R^{\prime}_{n}$ (where $R_n$ and $R^{\prime}_{n}$ are different random
numbers). The inclination relative to the line of sight is
calculated by considering the triangle formed by unit vectors along the
spin axis and the line of sight with respect to the cone axis. As 
the line of sight is at an angle $\alpha$ with respect to the cone axis, then
\begin{equation}
	\cos i_{n}  =  \sin\alpha \sin\theta_{n}\cos\phi_{n} +
	\cos\alpha \cos\theta_{n}\, .
\end{equation}
Since measurements of inclination derived from projected rotational
velocities cannot distinguish between $i$ and $\pi-i$, the
effective value of $\sin i$ is given by
\begin{equation}          
	\sin i_{n}  =  \sin
	(\cos^{-1}(\vert\sin\alpha \sin\theta_{n} \cos\phi_{n} +
	\cos\alpha \cos\theta_{n} \vert) )\, .	
\end{equation}

A Monte Carlo method is used to determine the expected distribution of
$\sin i$, for a given $\lambda$ and $\alpha$. 
A set of $\sin i$ values is evaluated for random values of
$\theta$ and $\phi$. The results are then ordered to define the
cumulative distribution function of $\sin i$. This cumulative  distribution
can then be used to determine a representative set of $\sin i$ values 
by generating random numbers between 0 and 1.
 
The right hand panels in Fig.~1 show the cumulative $\sin i$
probability distributions calculated for a uniform distribution and for
well aligned distributions with $\lambda = 25^{\circ}$ and various
values of $\alpha$. For $\lambda = 25^{\circ}$ the true
$\sin i$ distribution is significantly different from that of a uniform
distribution. Thus with sufficient $\sin i$ measurements it should be
possible to differentiate between well-aligned and 
uniform distributions, irrespective of the inclination $\alpha$.

Things become less clear when $\lambda$ is increased to, say,
$60^{\circ}$. In this case the results at low $(\alpha \simeq
20^{\circ})$ and high
$(\alpha \simeq 75^{\circ})$ inclinations 
could still be differentiated from a uniform
distribution. However, at intermediate inclinations the
$\sin i$ distribution becomes quite similar to that for a uniform
distribution. Thus a weakly aligned spin-axis distribution with large
$\lambda$ might only be discernible if its mean inclination is either
low or high.

Of course, these simple considerations have so far ignored the
alterations to the observed $\sin i$ distribution that are imposed by
selection effects in the data and by measurement uncertainties, which
are discussed in the next section.

\subsection{Measured ${\bf \sin i}$ distribution}
\label{actual spin axis}
An expression for the observed value of $\sin i$ is obtained from
equation~1 and by assuming that the stellar radius is proportional to
the product of its angular diameter and distance.
\begin{equation}
\sin i_{\rm obs}  =   k\,P_{\rm obs}  (v\sin i)_{\rm obs} / (A_{\rm obs}
D_{\rm est})\, ,
\end{equation}
where $k$ is a constant (appropriate to the units used) of
\begin{equation}
k = A_{\rm true} D_{\rm true} / (P_{\rm true} v_{\rm true})\, ,
\end{equation}
$P_{\rm obs}$ 
is the observed period of a star,
$(v\sin i)_{\rm obs}$ is its measured 
projected equatorial velocity, $A_{\rm obs}$ its
angular diameter (derived from a magnitude and colour via a
Barnes-Evans relation -- see section~\ref{barnesevans}) and $D_{\rm
  est}$ is the distance to the star estimated
by some independent method.

Uncertainties in the 
observed $v\sin i$, period, angular diameter and
distance estimate can be represented as Gaussian distributions with
normalised standard deviations of $\delta_p$, $\delta_v$, $\delta_A$ and
$\delta_D$, such that $P_{\rm obs} = P_{\rm true}(1 + \delta_P U)$ etc.
where $U$ is a random numbers drawn from a Gaussian distribution with
mean of zero and unit standard deviation. Hence we can write
\begin{equation}     
\sin i_{obs}  = \sin i_{true} \left(\frac{1 + \delta_{PV}U_1}{1 + \delta_{AD}U_2}\right)
\label{modelsini}
\end{equation}
where  $\delta_{PV} = \sqrt{\delta_{P}^{2}+\delta_{V}^{2}}$,
\ $\delta_{AD}=\sqrt{\delta_{A}^{2}+\delta_{D}^{2}}$ 

\noindent{and $U_1$ and $U_2$ are different random numbers with a mean of
zero and a standard deviation of 1. For the moment $\delta_{PV}$
and $\delta_{AD}$ can be considered as empirically derived
constants. Their values are discussed in section~\ref{errors}.}

In addition to measurement uncertainties there are
thresholds below which either period or $v \sin i$ cannot be
measured. These thresholds are equivalent to a lower limit to $\sin i$
below which rotational modulation and periodicity would not be detected,
and a resolution limit defining a threshold for $v\sin i$
detection. The latter is reasonably well defined from the observational
data, but there are significant astrophysical uncertainties in the
former -- e.g. the latitude distribution of spots on a young, active
star (see Jeffries 2007a for a discussion). We chose to represent this
observational bias as a simple cut-off value $\sin i_{\rm min}$, below
which a period could not be obtained for a star.  In Figure 2 we
show that the value of $\sin i_{\rm min}$ has a non-negligible effect on
the observed $\sin i$ distribution, so it is treated as a free
parameter in the following analysis and allowed to vary between zero and 
0.71.  That is to say we do not specify a lower limit of $\sin i$, and
at worst we expect to be able to measure the period and $v\sin i$ of stars 
with inclinations 45$^\circ$ and above. In section 5.1 it is shown that this
 is justified by the available observations.

To model the effects of the resolution limit for projected radial
velocity measurement, $(v\sin i)_{\rm min}$, we require an estimate of
the distribution of $v_{\rm true}$. The approach used here follows
\citet{Jeffries2007b} whereby the intrinsic $v_{\rm true}$ distribution
is represented as a combination of a uniform distribution and an
exponential decay. For the Monte Carlo analysis a fraction $\gamma$ of
velocities are drawn from a uniform distribution between zero and
$v_{\rm max}$ and the remainder from a cumulative exponential
distribution of the form $P(v) = \exp(-v/\beta)$. In practice
the $\sin i$ distribution is not sensitive to the exact form of the
$v_{\rm true}$ distribution so parameters defining the velocity
distribution ($\gamma, \beta, v_{max}$ and $v\sin i_{min}$) can be
estimated by matching the distribution of $(v \sin i)_{\rm obs}$.

To take into account the threshold values of $v\sin i$ and $\sin i$
in our simulations, any realisation with $\sin i_{true} < \sin i_{\rm
min}$ or $(v\sin i)_{obs}<(v\sin i)_{\rm min}$ is excluded from the
$\sin i$ distribution, since these
would not be present in an observed data set.

\begin{figure*}
\centering
\begin{minipage}[htb]{0.97\textwidth}
\includegraphics [width = 170mm]{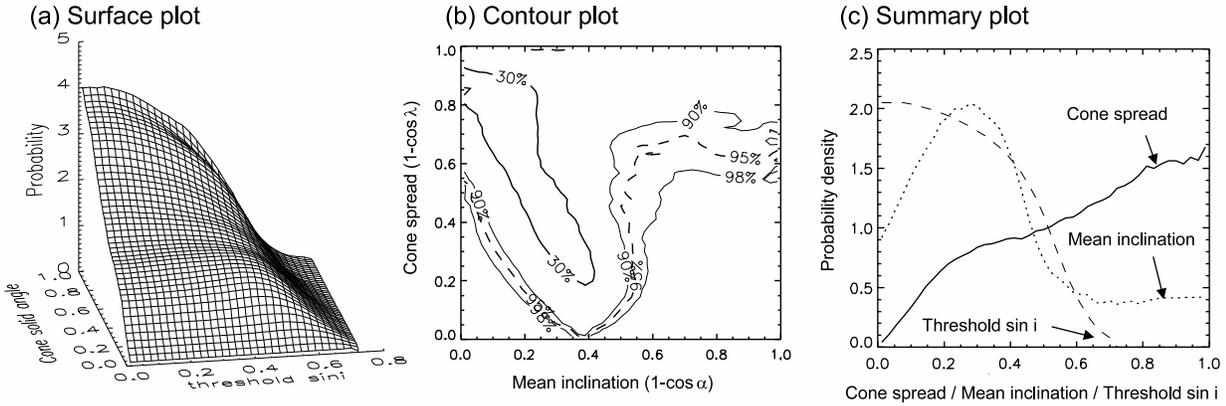}
\end{minipage}
\caption{Analysis of the $\sin i$ distribution measured on a group of
36 stars to determine the probability density of parameters describing
the underlying distribution of spin axes as a function of cone
spread (or solid angle) 
$(1-\cos (\lambda ))$ , mean inclination, $(1-\cos (\alpha ))$ and threshold 
in $\sin i$. The measured distribution was derived for stars in Pleiades 
(see Table 3 and section 4.4) }
\label{threeplot}
\end{figure*}

Figure 2 shows Monte Carlo simulations of the cumulative $\sin i$
distribution with typical levels of uncertainty: 10 per cent in
combined period and projected radial velocity and 10 per cent in
combined angular diameter and distance (see section 4.6). Results are
shown for three values of mean inclination, $\alpha=30^{\circ}$,
$60^{\circ}$ or $80^{\circ}$ and for $\sin i_{\rm min}=0.2$, 0.4 or
0.6. The solid lines show results for $\lambda=30^{\circ}$ and the
dashed line shows the probability density for a uniform distribution of
spin axes ($\lambda=90^{\circ}$).  Introducing uncertainties reduces
the difference in $\sin i$ distribution between the
$\lambda=30^{\circ}$ cone and the uniform distribution (compare with
Figure~1). However, the distributions are still quite different for
either low or high values of $\alpha$. For intermediate values
($45^{\circ}\le\alpha\le75^{\circ}$) it may still be possible to
resolve the difference provided $\sin i_{\rm min} \approx 0.2$. Above this,
the effect of increasing $\sin i_{\rm min}$ is to compress the
distributions along the $\sin i$ axis effectively erasing the
distinction between the aligned and random spin-axis distributions.

These results were calculated for a velocity distribution described by 
$\gamma=0.33$, $\beta=37$\,km\,s$^{-1}$, $v_{\rm max}=140$\,km\,s$^{-1}$ 
and $(v\sin i)_{\rm min} = 3.6$~km\,s$^{-1}$ (appropriate for the Pleiades -- see
section 4.5).  Also shown as a dotted line in Figure 2 are results
calculated assuming no lower cut off in $v\sin i$. This produces only small
changes in the modelled  $\sin i$ distributions, justifying the use of a
simple representation for the velocity distribution.

\subsection{Fitting parameters to measured distributions}
\label{fitting}
In our models there are three unknown parameters that define the
observed $\sin i$ distribution -- the opening half-angle of the cone,
$\lambda$, that describes the degree of spin-axis alignment, the mean
inclination $\alpha$ and $\sin i_{\min}$ which we will refer to as
$\tau$. The next step in the analysis is to determine how the $\sin i$
distribution of a group of stars can be analysed to determine the
underlying parameters $(\lambda, \alpha, \tau)$, of which we are most
interested in determining $\lambda$.

Elements of the measured $\sin i$ distribution are ordered to produce a
cumulative distribution function. This is then compared with a Monte
Carlo model $\sin i$ distribution using a Kolmogorov-Smirnov (K-S)
test. This gives an estimate of the probability that the measured data
could be drawn from the same distribution as the model data set. The
larger the estimated probability the more likely it is that the model
parameters ($\lambda$, $\alpha$, $\tau$) represent the measured data.

The K-S probability is calculated for all values of $\lambda_i,
\alpha_j, \tau_k$ and normalised by setting the integral over all
parameter space to be unity, to give the probability matrix
$\Delta^m(\lambda_i, \alpha_j, \tau_k)$. 
To visualise results it is useful to sum over one or two of the
independent variables to produce contour plots of probability density
for two parameters of interest or line plots for one parameter of
interest respectively.  Figure 3 shows typical results for a set of 36
stars in the Pleiades (see section 4). Figure~3a shows a surface plot of
the probability density as a function of ``cone spread'',
(or cone solid angle -- defined as $1-\cos\lambda$), and the $\sin i$ threshold, $\tau$. 
 These results show that there is a range of $\lambda$ and
$\tau$ that give reasonable fits to the measured data. Very roughly,
they indicate that the cone spread lies between $\approx$0.4 and 1 (corresponding to
$55^{\circ}<\lambda \leq 90^{\circ}$) and that $\tau < 0.5$. 
The modelling suggests then that the measured data are most likely drawn
from a nearly uniform distribution, but that there is still a finite
probability of a partially aligned distribution of spin axes.

To put this on a quantitative basis we can look at a a contour plot of
cone spread against $1-\cos\alpha$ (Fig.~3b).  The contours contain the
labelled percentage of the summed, normalised K-S probability. This
shows that if the cone inclination is small or close to unity ($\alpha$
close to $0^{\circ}$ or $90^{\circ}$) then the results could only be
consistent with a large cone spread $1 - \cos\lambda > 0.6$, or
$\lambda > 66^{\circ}$).  However, for intermediate values of
inclination the measurements are
consistent with almost any value of cone spread.  Figure 2 shows why
this is. At intermediate inclinations there is a much smaller
difference between the $\sin i$ distributions for well-aligned and
randomly orientated spin axes.

Figure 3c shows a summary of the results for the measured $\sin i$
distribution. The solid line shows the probability density of
cone spread $(1 -\cos\lambda)$. Integrating under this curve gives
a $90$ per cent probability that the cone solid angle is greater than
0.21, corresponding to a cone angle $\lambda > 38^{\circ}$. We can
also say with 90 per cent confidence that the threshold in $\sin i$ 
is less than 0.5.
 
\begin{figure*}
\centering
\begin{minipage}[b]{0.9\textwidth}
\includegraphics [width = 150mm]{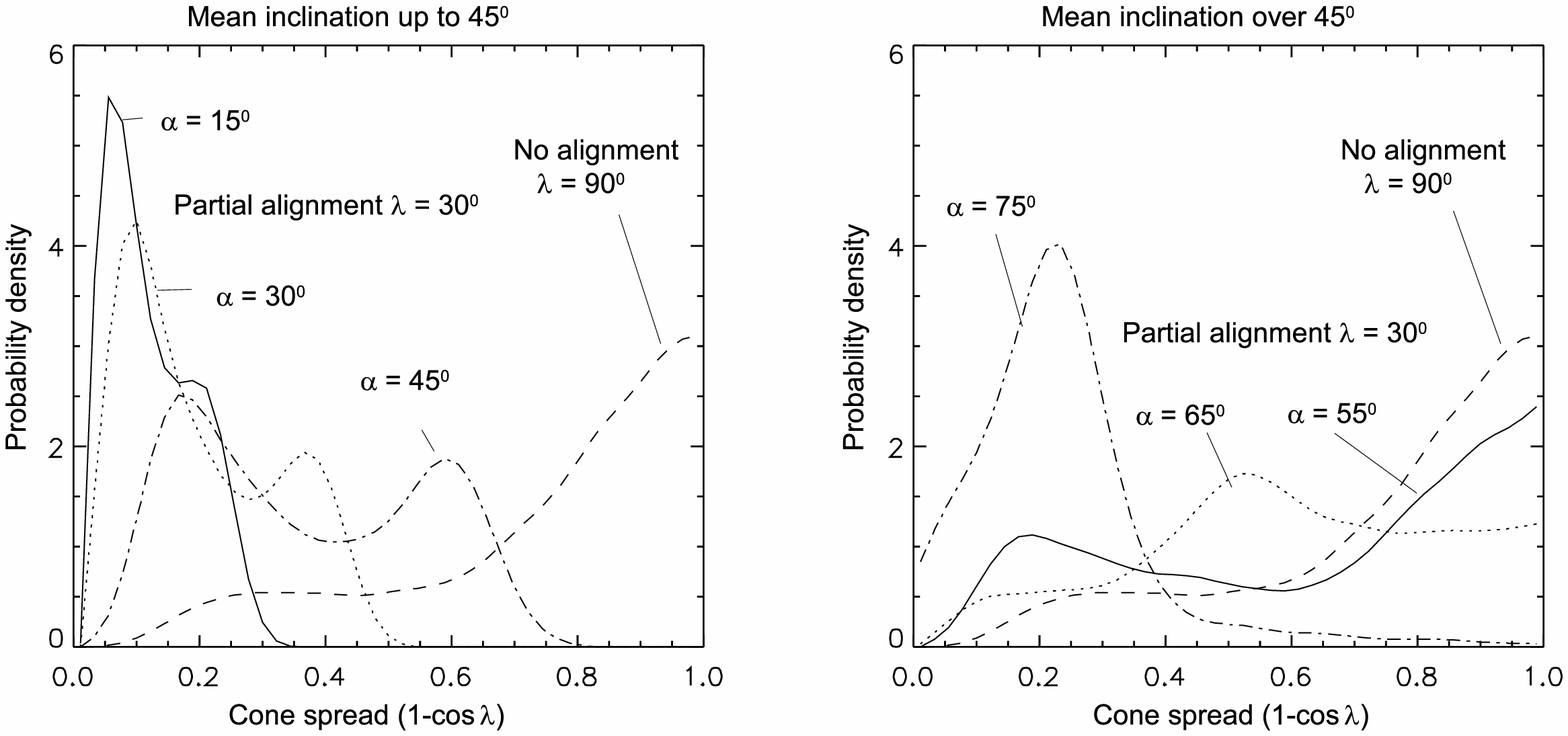}
\end{minipage}
\caption{The probability density as a function of cone spread $(1-\cos
\lambda )$ estimated from the analysis of the $\sin i$ distributions of
300 simulated stars. The curves show the results for
simulated datasets with partially aligned rotation axes ($\lambda =
30^{\circ}$) with different mean inclination angles, compared with a
simulated dataset with random spin-axis orientation
($\lambda=90^{\circ}$, dashed line). The left hand plot show the
comparison for cone inclinations, $\alpha = 15^{\circ}$, $30^{\circ}$
and $45^{\circ}$. The right hand plot shows the comparison for $\alpha
= 55^{\circ}$, $65^{\circ}$ and $75^{\circ}$.}
\label{effect uniform}
\end{figure*}

\subsection{Conditions for discrimination of cone angle}
The example in Figure 3 demonstrates how an observed $\sin i$
distribution could be analysed to investigate the underlying distribution
of spin-axis orientation for a group of stars in a cluster. 
This section considers how useful the method might be in practice. 
Specifically it considers:
\begin{enumerate}
\item Under what conditions will this method correctly recover
the underlying distribution of spin-axis orientation?
\item \indent What effect does sample size and uncertainty in 
individual $\sin i$ measurements have on the accuracy of the method?
\end{enumerate}
To address the first question the Monte Carlo method is used to
generate $\sin i$ distributions for two scenarios. The first represents
the case where there is significant alignment of spin axes of stars in
the cluster corresponding to $\lambda=30^{\circ}$. The second case
represents a uniform distribution with $\lambda=90^{\circ}$. 
The $\sin i$ values are produced
for sets of 300 stars with typical levels of uncertainty in $(\sin
i)_{\rm obs}$,
namely $\delta_{PV} = \delta_{AD}=10$ per cent, a representative
$v_{\rm true}$ distribution and $\tau = 0.4$.  The cumulative $\sin i$ 
distributions are then analysed using the method described above to
determine the probability density of cone spread, which should
reflect the underlying distribution of spin axes.

The results are shown in Fig.4 for increasing values of $\alpha$. For
low values of $\alpha$ ($<45^{\circ}$) the distribution is peaked
towards low values of $1- \cos \lambda$, roughly corresponding to the
input value of $\lambda$. At high values of $\alpha$ ($\geq
75^{\circ}$) this is also the case. However at intermediate values, the
probability density becomes more uniform indicating that the $\sin i$
distribution could have arisen from almost any value of $\lambda$. The
reasons for this were discussed in section 3.1 and need not be repeated here.

Where the probability density falls to zero at one end or other of the
distribution, we can calculate limits on $\lambda$. For example,
integration under the dashed line in Fig. 4 (where the input value of
$\lambda$ was $90^{\circ}$) returns a 90 per cent confidence limit that
$\lambda > 50^{\circ}$. This limit is the figure of merit we choose to
characterise a $\sin i$ distribution, indicating whether it results
from an aligned distribution or not.

\noindent
\begin{table}
\caption{Lower limit to the cone angle, $\lambda$, found from input simulated datasets with
  random spin-axis orientation.}
\begin{tabular}{lccc}
\hline 
No. stars  & Threshold in& \multicolumn{2}{c}{90 per cent lower limit} \\
analysed &  $\sin i$, $\tau$ & \multicolumn{2}{c}{of cone angle, $\lambda$ (deg)} \\ \hline 
 Uncertainties,& $\delta_{PV}$ \& $\delta_{AD}$  & 0.10 \& 0.10 & 0.15 \& 0.15 \\ \hline
 10 & $\sin i>0.2$ & $>30\pm$6 & $>29\pm$7 \\ 
 & $\sin i>0.4$ & $>30\pm$4 & $>29\pm$5 \\ \hline 
30 & $\sin i>0.2$ & $>39\pm$4 & $>33\pm$9 \\ 
 & $\sin i>0.4$ & $>36\pm$6 & $>33\pm$5 \\ \hline 
100 & $\sin i>0.2$ & $>46\pm$6 & $>40\pm$7 \\ 
 & $\sin i>0.4$ & $>47\pm$2 & $>48\pm$3 \\ \hline 
300 & $\sin i>0.2$ & $>47\pm$3 & $>48\pm$5 \\  
 & $\sin i> 0.4$ & $>49\pm$5 & $>44\pm$5 \\ \hline 
\end{tabular}
\label{table1}
\end{table}

To determine the effect of sample size, measurement uncertainties and
the threshold $\tau$, a range of Monte Carlo simulations were
performed. These were divided into runs with $\lambda=90^{\circ}$ and
$\lambda=30^{\circ}$. The results of analysing the $\sin i$
distributions are listed in Tables 1 and 2.

The main conclusions from these simulations are:
\begin{enumerate}

\item For an input $\lambda=90^{\circ}$ (Table 1), the lower limit that can be
placed on the recovered $\lambda$ value is quite insensitive to the
sample size. Observing many hundreds of stars does not give much
improvement over sample sizes of $\sim 30$. Neither are the results
sensitive to the actual value of $\tau$ or the exact value of the
measurement uncertainties.

\item For a strongly aligned distribution (input $\lambda=30^{\circ}$,
Table 2) we find that upper limits to $\lambda$ are only obtained if
the mean inclination $\alpha \leq 45^{\circ}$ or $\alpha \geq
75^{\circ}$. Again, the gains to be made by observing very large
samples of stars are not very significant, although samples of $\sim
100$ may be required to identify a strongly aligned spin-axis
distribution when $\alpha$ is large.
\end{enumerate}

\section{Measured \mbox{$\sin i$} distributions}
In this section we use the previously developed ideas to model the
observed $\sin i$ distributions obtained from published measurements
for stars in the young Pleiades and Alpha~Per open clusters.  
These clusters have ages of $\simeq 120$\,Myr and $\simeq 80$\,Myr
respectively, so many of the G- and K-stars are rotating fast enough to
produce measurable rotational broadening of their spectral lines --
giving $(v \sin i)_{\rm obs}$ -- and produce rotational modulation caused by
magnetic starspots -- giving $P_{\rm obs}$. Angular diameters are determined from reported 
$V$ and $K$ magnitudes using a surface brightness relation calibrated by interferometry
(Kervella et al. 2004).
\subsection{Measured data for the Pleiades and Alpha-Per clusters}

Even in the well-studied Pleiades and Alpha-Per clusters, there are
surprisingly few stars where $v \sin i$ and $P$ are both
known. The database for galactic open clusters (Mermilliod 1995)
currently lists 296 stars in the Pleiades cluster with known
$v \sin i$, and 56 with known $P$, not all of which
overlap. Similarly, there are 247 stars in Alpha-Per with
known $v\sin i$ and 66 with known $P$. Table 3 shows the
$(v \sin i)_{\rm obs}$ and $P_{\rm obs}$ for 44 stars in the Pleiades and 38
in Alpha~Per that have been identified as cluster members and for which
this simultaneous information is available.

Table 3 also lists the available stellar photometry, which is
subsequently used to check for unresolved binarity and estimate the
angular diameters. We tabulate a mean $V$ magnitude and the
(peak-to-peak) amplitude of the light curve modulation used to find the
stellar rotation period. The apparent $K$ magnitudes are
taken from the 2MASS catalogue (Cutri et al. 2003). A small
offset is applied to convert these to the CIT photometric system used in the
evaluation of angular diameter (see Section~\ref{barnesevans}) 
$K_{CIT} = K_{2MASS} + 0.024$, \citet{Carpenter2001a}.

\subsection{Estimation of angular diameter}					 
\label{barnesevans}

The method used to estimate stellar angular diameter in equation~9
follows that of O'Dell, Hendry \& Collier Cameron (1994).  A
Barnes-Evans relationship (e.g. Barnes \& Evans 1976) determines the
angular diameter of a star from its measured apparent magnitude and a
colour index. Whilst the approach of O'Dell et al. is appropriate for
the current application, the calibration data they used for the
Barnes-Evans relationship has been superseded. In addition, O'Dell et
al. used the $B-V$ colour index, but this is now known to be a
systematically unreliable temperature indicator in magnetically active,
spotted cool stars (e.g. Stauffer et al. 2003). We prefer to use $V-K$,
which is potentially more precise and appears less affected by stellar
activity and metallicity -- the latter being a potential source of
uncertainty in any calibration sample.

\noindent
\begin{table}
\caption{Upper limit to the cone angle, $\lambda$ found from input simulated datasets 
with cone angle $\lambda = 30^{\circ}$  and  $\tau=0.4$}
\begin{tabular}{lccc}
\hline 
No. stars  & Mean & \multicolumn{2}{c}{90 per cent upper limit} \\
analysed & Inclination & \multicolumn{2}{c}{to cone angle, $\lambda$ (deg)} \\ \hline 
 Uncertainties, & $\delta_{PV}$ \& $\delta_{AD}$  & 0.1 \& 0.1 & 0.15 \& 0.15 \\ \hline
 10 & $15^{\circ}$ & $<41\pm$4 & $<52\pm$8 \\ 
 & $30^{\circ}$ & $<64\pm$12 & $<65\pm$8 \\  
 & $45^{\circ}$ & $<74\pm$6 & $<76\pm$10 \\ 
 & $75^{\circ}$ & \multicolumn{2}{c}{not resolved} \\ \hline
30 & $15^{\circ}$ & $<39\pm$6 & $<42\pm$4 \\  
 & $30^{\circ}$ & $<53\pm$6 & $<54\pm$3 \\  
 & $45^{\circ}$ & $<72\pm$6 & $<73\pm$11 \\ 
 & $75^{\circ}$ & $<74\pm$6 & $<80\pm$3 \\ \hline 
100 & $15^{\circ}$ & $<32\pm$1 & $<34\pm$5 \\  
 & $30^{\circ}$ & $<51\pm$4 & $<58\pm$4 \\
 & $45^{\circ}$ & $<64\pm$3 & $<68\pm$2 \\ 
 & $75^{\circ}$ & $<66\pm$14 & $<56\pm$16 \\  \hline  
300 & $15^{\circ}$ & $<30\pm$3 & $<30\pm$5 \\
 & $30^{\circ}$ & $<49\pm$1 & $<47\pm$6 \\ 
  & $45^{\circ}$ & $<66\pm$2 & $<67\pm$3 \\ 
  & $75^{\circ}$ & $<53\pm$17 & $<53\pm$11 \\ \hline  
  
\end{tabular}
\label{table2}
\end{table}

\noindent
\begin{table*}
\caption{ Estimated values of $\sin i$ for stars in the Pleiades and Alpha Per clusters using; the
Barnes-Evans relation in equation 11, a mean distance of
131.8$\pm$2.4pc and a colour excess of 0.032 for the Pleiades and a 
distance of 176.2$\pm$4.9pc and a colour excess of 0.10 for the Alpha Per cluster. 
The full table is available at Blackwell Synergy as supplementary material
to the on-line version of this table.}
\begin{tabular}{lcccccccl} \hline 
Star & Period & $V\sin i$ & Apparent & Apparent & Colour & Variation &
Estimated & Estimate \\ Name & with ref. & with ref. & Magnitude &
Magnitude & excess & magnitude & diameter & of $\sin i$ \\ ~ & (days) &
(km/s) & $V_{mean}$ & $K_{2MASS}$ & $V-K_{CIT}$ & ${\Delta V}$ & A
(arcsec) \\ \hline
H\sc{ii} 191 & 3.100~a & ~~9.1~f & 14.38 & 10.61 & 3.75 & 0.04 & 0.047 & 0.84   \\
H\sc{ii} 253 & 1.721~b & ~38.2~f & 10.66 & ~8.95 & 1.69 & 0.12 & 0.072 & 1.27   \\
H\sc{ii} 263 & 4.820~a & ~~7.8~f & 11.63 & ~9.39 & 2.22 & 0.16 & 0.065 & 0.81   \\
H\sc{ii} 293 & 4.200~c & ~~5.7~f & 10.79 & ~9.06 & 1.71 & 0.02 & 0.067 & 0.50   \\
H\sc{ii} 314 & 1.479~d & ~41.9~f & 10.56 & ~8.90 & 1.64 & 0.09 & 0.073 & 1.19   \\
H\sc{ii} 320 & 4.600~c & ~10.8~f & 11.04 & ~8.87 & 2.15 & 0.06 & 0.080 & 0.87   \\
H\sc{ii} 345 & 0.723~e & ~18.9~f & 11.40 & ~9.27 & 2.11 & 0.07 & 0.066 & 0.29   \\
H\sc{ii} 357 & 3.400~c & ~10.0~f & 13.32 & 10.02 & 3.28 & 0.07 & 0.057 & 0.83   \\
H\sc{ii} 739 & 0.917~e & ~14.4~f & ~9.44 & ~7.94 & 1.48 & 0.03 & 0.108 & 0.17 **  \\
H\sc{ii} 883 & 7.200~a & ~~3.8~f & 13.05 & 10.25 & 2.78 & 0.10 & 0.048 & 0.80   \\
 \hline  
\multicolumn{9}{l}{** Stars identified as probable binaries and
excluded from the analysis of the $\sin i$ distribution }\\\\
\multicolumn{9}{l}{The letter following the period
value denotes the reference for the period and photometric data.}\\
\multicolumn{9}{l}{The letter following the $V\sin i$ value indicates
its source. The sources are as follows: a
\citet{Krishnamurthi1998a},}\\
\multicolumn{9}{l}{ b \citet{Marilli1997a}, c \citet{Prosser1995a}, d
\citet{Prosser1993a}, e \citet{Messina2001a}, f \citet{Queloz1998a},}\\
\multicolumn{9}{l}{ h \citet{Stauffer1987a}, i \citet{Soderblom1993a},
j \citet{Stauffer1989a}, k \citet{Stauffer1985a}} \\
\multicolumn{9}{l}{l \citet{Prosser1993a}, m \citet{Prosser1997a}, n
\citet{Bouvier1996a}, p \citet{ODell1994a}}\\
\end{tabular}
\label{table3}
\end{table*}

Kervella et al. (2004) provide a recalibration of the Barnes-Evans 
relationship based on angular diameters
of main sequence and sub-giant stars measured by interferometry. 
They give the following relationship between angular diameter, $A$, (in units 
of arcseconds) de-reddened magnitude, $K_o$ and colour $(V-K)_o$

\begin{eqnarray}  
\log(A)  & = & (0.5170\pm 0.0017) + (0.0755\pm 0.0008)(V-K)_o \nonumber\\
         &   & - 0.2K_o\, .
\end{eqnarray}
This is valid for dwarfs of spectral type A0--M2 and has an
exceptionally small intrinsic dispersion ($<1\%$), although in our case
there are other sources of uncertainty in the estimated angular
diameter, notably uncertainty in apparent magnitude and interstellar
extinction. To allow for interstellar extinction the corrections given
by Rieke and Lebofsky (1985) ($A_V = 3.1E(B-V)$ and
$E(V-K)=2.74E(B-V)$) are applied to the above expression to give
\begin{eqnarray}  
\log(A)  & = & 0.5170 + 0.0755(V-K) - 0.2K \nonumber \\
         &   & -0.136E(B-V)\, ,
\end{eqnarray}
where $V$ is the peak apparent magnitude for the variable star
($V_{mean}-\Delta V/2$) (see Table 3). The rationale here is to
use the brightest value of $V$ to represent an unspotted photosphere. 
A mean value of colour excess, 
E(B-V) is used for each cluster.
A value of 0.032 is taken from An, Terndrup \& Pinsonneault(2007)
for the Pleiades and a value of 0.10 taken 
from \citet{Pinsonneault1998a} for Alpha Per. The expression 
for $\log (A)$ is quite insensitive to the magnitude of
the colour excess, a very conservative uncertainty of 0.05 in $E_{B-V}$ would only 
change an estimated angular diameter by 1.5 per cent.

\subsection{Effects of binarity}
The stars in Tables 3 will inevitably include some unresolved
binaries. As the effect of binarity will be to decrease the apparent
magnitude (by up to 0.75 mag) and possibly redden the star with respect
to the intrinsic colour and magnitude of a single star, equation 11
shows that unrecognised binarity could change the estimated angular
diameter and deduced value of $\sin i$.

\begin{figure*}
\centering
\begin{minipage}[tb]{0.95\textwidth}
\includegraphics [width = 150mm]{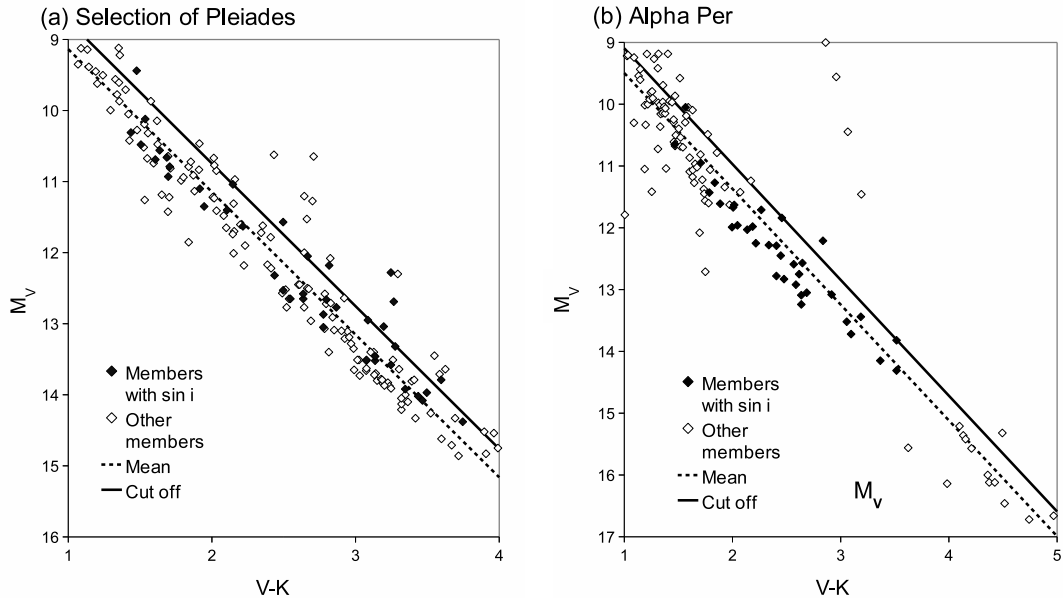}
\end{minipage}
\caption{(a) Selection of probable binaries from stars of known period
and $v\sin i$ in (a) the Pleiades and (b) Alpha Per. The solid line is located 0.4 
magnitudes above the dashed regression line through all cluster members.
Stars lying above the solid line are considered as probable binaries 
and excluded from the measured $\sin i$ distribution.}
\end{figure*}

To mitigate this, probable
binaries were eliminated by rejecting stars that lie
significantly above the mean cluster sequence in a colour magnitude
diagram. To do this in a systematic manner, reference stars identified
as being cluster members in the WEBDA data base which showed 80\% or
greater probability of membership from proper motion measurements were
plotted in a $V$ versus $V-K$ diagram (see Fig.~5).
Stars lying more than 0.4 magnitudes above the linear regression line
through these reference data were identified as probable binaries. When
this criteria is applied to the Pleiades data in Table 3 then 8/44 of
the original sample were eliminated. A similar analysis for Alpha Per eliminated
2/38 of the stars in Table 3.

\subsection{Cluster distances}
To calculate $\sin i$ from equation~7 we need cluster distances.  We
adopt $131.8\pm 2.4$\,pc (distance modulus 5.60\,mag) for Pleiades and
$176.2\pm 4.9$\,pc (distance modulus 6.23\,mag) for Alpha-Per, based on
multi-colour main sequence fitting (see Table~1 of
\citet{Pinsonneault1998a}). There has been controversy over the Pleiades
distance following a much lower parallax-based measurement using the
Hipparcos satellite of $120.2\pm 1.9$\,pc \citep{vanLeeuwen1997a,
vanLeeuwen2009a}.  The higher distance, consistent with the results of
main-sequence fitting and the distance to binary stars, has generally
been been adopted in the literature. These include: a distance modulus
of $5.60\pm 0.07$\,mag from radial velocity and interferometric
measurements of Atlas (HD 23850) \citep{Zwahlen2004a};
\citet{Munari2004a} used radial velocity and photometry of HD 23642 to
get a distance modulus of $5.60\pm 0.03$\,mag, although re-analysis of
the same data by \citet{Southworth2005a} gave a distance modulus of
$5.72\pm 0.06$\,mag; parallax measurements on three stars in Pleiades
made using the fine guidance sensors on the Hubble Space Telescope
\citep{Soderblom2005a} gave a distance modulus of $5.65\pm 0.02$\,mag.
The source of the discrepancy between the Hipparcos results and other
measurement techniques remains unresolved.  The distance to the
Alpha-Per cluster has not been investigated to the same extent as the
Pleiades. The Hipparcos distance of $172.4\pm 2.7$\,pc agrees
well with the main sequence fitting distance in this case.

\subsection{Distributions of projected velocity}
For the Monte-Carlo analysis we need an estimate of the intrinsic
properties of the equatorial velocity distribution (see section 3.2).
These are estimated from the observed $v \sin i$ data as follows.  The
maximum equatorial velocity, $v_{\rm max}$, and minimum projected
velocity, $v\sin i_{\rm min}$ are taken as the maximum and minimum
observed $v\sin i$ values. The true distribution of equatorial
velocities is parameterised in terms of $\gamma$ and $\beta$ (see
section 3.2). These coefficients are chosen to maximise the probability
that a model $v \sin i$ distribution and the observed data set are
drawn from a common distribution using a K-S test. Values of $\gamma =
0.33$ and $\beta = 37$\,km\,s$^{-1}$ for Pleiades give a K-S
probability of 0.8. Values of $\gamma = 0.16$ and $\beta =
54$\,km\,s$^{-1}$ for Alpha~Per give a probability of 0.7. We
re-emphasise that the exact choice of these parameters has very little
effect on our Monte-Carlo models of the observed $\sin i$ distribution
(see section 3.2 and Fig.~2).

\subsection{Measurement uncertainties}
\label{errors}

To model the measured $\sin i$ distribution we assume that
statistical variations in the parameters used to derive $\sin i$ are
Gaussian distributions  (see equation~\ref{modelsini}) with normalised
uncertainties that are derived from the measured data and published
uncertainties. The source papers give 
uncertainties for about half the periods in Tables 3. The
average of this subset of measurements gives
$\delta_P = 0.03$. The average normalised
uncertainty for the velocities from Queloz el al. (2001) in Table 3 is
$\delta_{V} = 0.10$. Including uncertainties for data from other sources increases the
average to 0.12. On the basis of these results the normalised
uncertainty due to the combined effects of period and velocity,
$\delta_{PV}$, is taken to be either 0.10 or 0.15.

The main source of uncertainty in estimated angular diameters is
uncertainties in apparent magnitudes. Normally these should be
relatively low, say $\pm 0.05$~mag for $V$ and even less for $K$.  However,
there is additional uncertainty for the stars used here which exhibit
rotational modulation.  From Tables 3 the rms magnitude
variation is $\Delta V_{\rm rms} = 0.09$~mag.  Taking this as an
upper bound for the uncertainty in both $V$ and $K$ then, from
equation~16, the uncertainty in angular diameter is $\pm 5$ per cent.
If we include the small intrinsic uncertainties in the Barnes-Evans
relationship, $<1$ per cent (Kervella et al. 2004) and colour excess
this gives a total uncertainty of about 6 per cent in the estimated
radii.

The uncertainty in distance can be considered in two parts. First there
is a random uncertainty due to the location of a star within the
cluster. This depends on the radial depth of the cluster. We assume
that the radial depth of the sample is similar to its tangential width
on the sky. The Pleiades sample is located a mean distance of
$0.8^{\circ}$ from the cluster centre, corresponding to a normalised
distance uncertainty of 0.014. For Alpha~Per the sample is scatter over
a larger area, corresponding to a normalised distance uncertainty of 0.031
On this basis a value of $\delta _{AD}=0.10$ gives a
conservative estimate of the combined effects of uncertainty in angular
diameter and distance.

In addition to the uncertainties applied to individual stars, there is
an uncertainty, $\delta_{D{\rm est}}$, in the estimate of average distance to
the cluster. This produces a systematic
error in the analysis, which becomes significant if the normalised
uncertainty in the distance, $D_{\rm est}$ is comparable with the
uncertainty in the {\it mean} value of $\sin i$ due to the combined effects
of the other sources of error measured in $N$ stars.
\begin{equation}
\delta_{D{\rm est}} \sim \sqrt{(\delta^2_{PV}+\delta^2_{AD})/N}
\end{equation}	 
Thus for $N=36$ stars, with $\delta_{PV}$and $\delta_{AD}$ of 10 per cent, the
uncertainty in estimated distance becomes significant at a level of 2.4
per cent. As the assumed distance uncertainties to the Pleiades and
Alpha~Per are 1.8 and 2.8 per cent respectively, then the sensitivity
of our results to variations in this value must also be considered.

\begin{figure*}
\centering
\begin{minipage}[tb]{0.9\textwidth}
\includegraphics [width = 150mm]{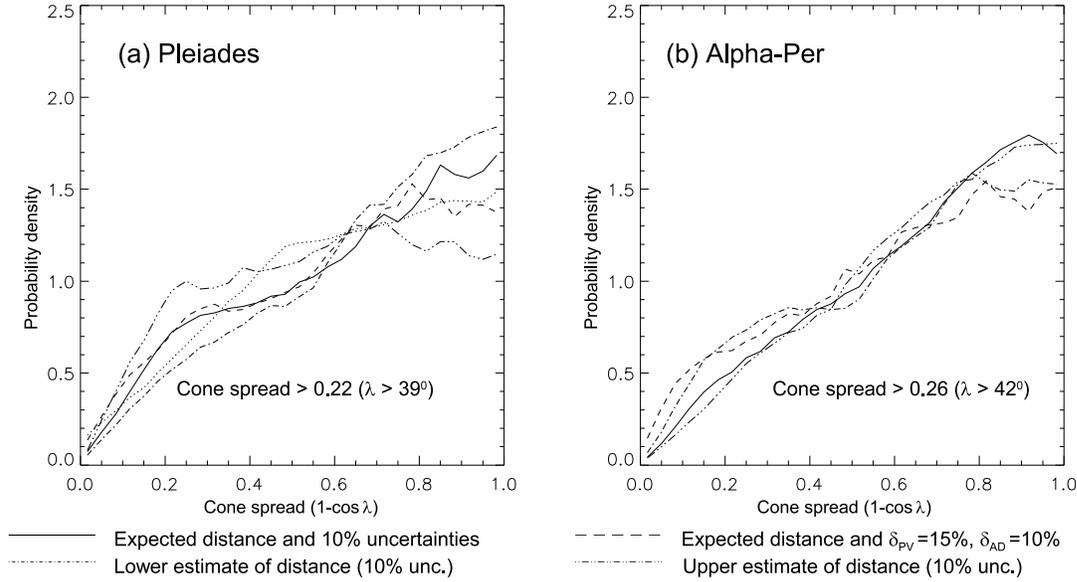}
\end{minipage}
\caption{Analysis of measured $\sin i$ distributions for the Pleiades and
Alpha~Per. The solid line shows the probability density of cone spread
$(1 - \cos \lambda)$
that describes the underlying distribution of spin axes for normalised
uncertainties of $\delta _{PV}=\delta_{AD}=0.10$. The dashed line shows the
effect of increasing the normalised errors to 0.15 and 0.10. The
dashed/dotted curves show the effects of varying the assumed distance to the
cluster by plus or minus one standard deviation. The dotted line in the left 
hand plot shows the probability density of cone spread calculated 
assuming the a lower estimate of distance based on Hippacos data.}
\label{results}
\end{figure*}

\section{Results}

\subsection{Probability density of cone angle for Pleiades and Alpha~Per}

To investigate the degree of spin-axis alignment we look at the
constraints we can place on the cone angle $\lambda$.
The Pleiades stars, filtered for binaries, 
were modelled using the method described in section 3.
Normalised uncertainties of $\delta_{PV}= \delta_{AD}=0.10$ 
were initially assumed. Figure 3 showed plots of the probability 
distributions corresponding to
$\lambda$, mean inclination $\alpha$, and threshold $\sin i$ value
$\tau$.  Figure 6 now shows a similar plot of the probability density
of cone spread ( $1 - \cos \lambda$) for both Pleiades and
Alpha~Per, evaluated for different levels of combined uncertainty in period and $v\sin i$ 
and for the cluster distances fixed at the assumed value plus or minus their formal error bars.

For both clusters there is a clear increase in probability for larger
values of cone spread $(1 - \cos \lambda)$ and therefore of $\lambda$
itself.  A qualitative comparison with the simulations shown in Fig.~4
demonstrate that this behaviour suggests a random spin-axis
orientation, rather than one with a small $\lambda$.  By integrating
the probability distributions we can say with 90 per cent confidence
that $\lambda > 39^{\circ}$ for the Pleiades and $\lambda > 42^{\circ}$
for Alpha Per (see Table 5).  The most likely model for both clusters
is one with close-to random spin-axis alignment ($\lambda \simeq
90^{\circ}$).  Figure~3b shows that a lower value of $\lambda$ would
have to be ``disguised'' in the data by an intermediate value of
alignment, $\alpha$ (see the discussion in section 3.3).

\noindent
\begin{table}
\caption{Constraints on the cone angle, $\lambda$, and on the threshold
  value of $\sin i$, $\tau$ for the Pleiades and
Alpha Per for different assumed distances and levels of measurement uncertainty.}
\begin{tabular}{llccc}
\hline
Distance & Uncertainty &Cone Angle $\lambda$ &Threshold
 $\tau$\\
 to cluster & $\delta_{PV}$ , $\delta _{AD}$ & 90\% limit&
 90\% limit \\ \hline
 Pleiades& & & \\
 131.8 pc &  0.10, 0.10 & $>39^{\circ}$ & $<0.48$ \\ 
 131.8 pc & 0.15, 0.10   & $>37^{\circ}$  & $<0.50$\\ 
 129.4pc &  0.10, 0.10  & $>43^{\circ}$  & $<0.48$\\ 
 134.2 pc &  0.10, 0.10  & $>36^{\circ}$  & $<0.47$\\ \hline
 \multicolumn{4}{l}{Pleiades using Hipparcos distance} \\
 120.2 pc &  0.10, 0.10 & $>40^{\circ}$ & $<0.63$ \\ \hline
Alpha Per & & & \\
 176.2 pc &  0.10, 0.10  & $>42^{\circ}$  & $<0.52$\\ 
 176.2 pc & 0.15, 0.10   & $>37^{\circ}$ & $<0.57$\\ 
 171.3 pc &  0.10, 0.10  & $>39^{\circ}$  & $<0.51$\\ 
 181.1 pc &  0.10, 0.10  & $>44^{\circ}$  & $<0.54$\\ \hline
\end{tabular}
\label{table5}
\end{table}

For both the Pleiades and Alpha Per we also find similar 90 per cent
upper limits to $\tau$ of $\simeq 0.5$, suggesting that it is not until
they have inclinations below $30^{\circ}$ that stars become unlikely to
exhibit significant rotational modulation in their light curves. This
supports the assumption made in section 3.2 that  $\tau$ will always be 
less than 0.71, which corresponds to $i_{min} \leq 45^{\circ}$. 
Figure~6 and Table 4 also illustrate how robust these results are to
alterations in our assumptions about the level of measurement
uncertainty. Increasing the
$\delta_{PV}$ to 0.15 does not change the probability distributions 
in any significant way and has only a modest impact on the 
90 per cent lower and upper limits to $\lambda$ and $\tau$.

Changing the assumed distance could have more influence. In Fig.6 we
show curves derived for assumed distances that are equal to the main
sequence fitting distances plus or minus an error bar. For the Pleiades reducing the 
distance would yield a larger $\lambda$ lower limit, whereas increasing the distance
hints at the {\it possibility} of some alignment though still
consistent with a uniform distribution. We could discard the main-sequence 
fitting distance entirely and instead adopt the lower Hipparcos distance. This case
is also shown in Fig.6. Whilst the cone angle probability becomes 
a little more uniform there is still evidence of alignment. The results in Table 4
show that $\lambda$ is still greater than $40^{\circ}$. We caution the reader that this should not
be taken as evidence that these methods are insensitive to the assumed distance. 
As discussed in section 4.6, a distance accurate to a few per cent is really require
 for trustworthy constraints on $\lambda$.

\subsection{New distance estimates to the Pleiades and Alpha~Per}

If we adopt the assumption that the spin-axes really are randomly
oriented then we can derive an independent distance(O'Dell
et al. 1994). This provides a first check for possible alignment 
of spin axes since if the distance derived assuming random orientation does not
agree with the established distance there is a strong indication of
partial alignment which can be investigated using the method 
described above. Note the converse is not true since a matching distance
can also be produced by a partially aligned cluster at a favorable
average inclination ($45^{\circ}$ to $75^{\circ}$).

Combining the $v \sin i$ and period data in Table 3 gives
an estimated distance to the Pleiades  of $133\pm 7$pc. 
This is in good agreement with the value found 
from main sequence fitting of $131.8\pm 2.4$\,pc adopted in the last section
\citep{Pinsonneault1998a},  but significantly higher than
the Hipparcos value of $120.2\pm 1.9$\,pc \citep{vanLeeuwen2009a}. The
uncertainty in our independent distance estimate is too large to completely rule
out the Hipparcos result, but clearly favors the more
conventional main-sequence fitting measurement (assuming that spin axes
truly have a random orientation!). 

For Alpha~Per the data in Table 3 gives an estimated distance of
$182\pm 11$pc. This agrees well with the distance given by main
sequence fitting, $176.2\pm 4.9$\,pc, and within one standard deviation of the value derived 
from the new reduction of Hipparcos data, $172.6\pm 2.8$\,pc \citep{vanLeeuwen2009a}.

It is also interesting to compare these results with those of O'Dell et
al. (1994) who used a similar technique to find the distance to these
clusters, but with smaller samples. They found distances of $132 \pm
10$\,pc and $186 \pm 12$\,pc for the Pleiades and Alpha~Per
respectively. To some extent the agreement here is fortuitous.  O'Dell
et al. used an older Barnes-Evans calibration based on $B-V$ data and
did not exclude probable binaries.  Using our data set with the surface
brightness $B-V$ relations of Kervella et al. (2004) would tend to
increase the estimated cluster distance, whereas our exclusion of
possible binaries reduces it.

\subsection{Biases in estimated cluster distances}

If the distribution of spin-axis orientation were not random then there
may be a systematic error in the distances estimated in the last
section and it might be possible to obtain agreement with the Hipparcos
distance for the Pleiades. In Fig. 7 we show the results of Monte-Carlo
simulations which investigate by what factor we would overestimate the
true distance for the cases of two aligned distributions with
$\lambda=15^{\circ}$ and $\lambda=45^{\circ}$ respectively, if we were
to analyse the $\sin i$ distributions under the assumption of random
alignment. In each case we have assumed that the mean inclination
$\alpha$ is randomly distributed between a maximum value of
$90^{\circ}$ and a minimum value of $15^{\circ}$ which corresponds to
spin-axes which almost point towards the observer. We take $15^{\circ}$
as a practical minimum value because for lower values we simply
wouldn't be able to measure any $v \sin i$ values.

The mean x-axis values for the two distributions shown in Fig.7 are
both 0.99. That is, the distance we would estimate (on average) is not
significantly biased by the assumption of random spin-axis alignment.
However, the widths of these distributions imply a significant
additional scatter. Clusters in which the spin-axes were highly aligned
($\lambda=15^{\circ}$), with a mean inclination $\alpha=15^{\circ}$
would have their distances under-estimated by a factor of two. If on
the other hand the mean inclination were to take its maximum value of
$\alpha=90^{\circ}$, the distance would be overestimated by a maximum
factor of about 1.2. 

If $\lambda$ is not $90^{\circ}$, 
then it seems reasonable to take a 68
per cent confidence interval about the mean of the $\lambda=15^{\circ}$
distribution in Fig. 7 as an indication of the maximum additional systematic
error in the estimated distance that is introduced by {\it assuming}
random spin-axis orientation. Quantitatively, this amounts to an additional 
uncertainty of $^{+18}_{-32}$~percent.

\begin{figure}
\centering
\begin{minipage}[b]{0.85\textwidth}
\includegraphics [width = 80mm]{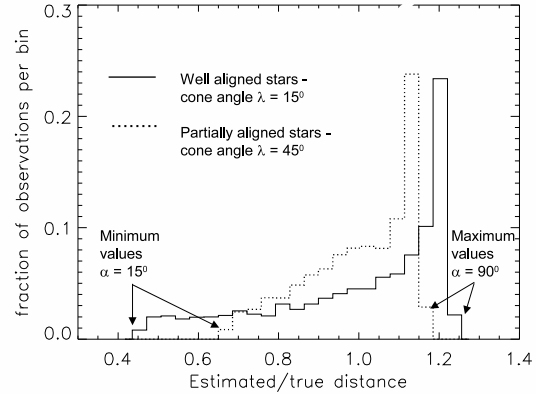}
\end{minipage}
\caption{Possible bias in the estimated distance to a cluster for well 
($\lambda = 15^{\circ}$) and partially aligned ($\lambda = 45^{\circ}$) 
stars in a cluster. The histogram shows the frequency of estimates 
of distance (normalized to the true distance) assuming that the mean axis 
of alignment of the cluster is oriented randomly in space.}
\label{bias}
\end{figure}

\section{Discussion and Summary}
The results in this paper show that analysis of the measured $\sin i$ values
for groups of as few as 36 stars in a cluster can be used to
investigate whether their spin axes are randomly oriented or 
well aligned in space. The information that can be gained is incomplete because
(a) information is lost in the measurement technique, principally in the
azimuthal direction of the spin axes and (b) there are significant
uncertainties in the measurements of parameters used to determine $\sin
i$, which blur its distribution function. 

The simulations in section 3.4 and Fig.~4 indicate that a cluster of
stars with random spin-axis orientation can be identified from the
strong increase in probability density with cone angle. In some cases
it is easy to distinguish this from a well-aligned distribution of spin
axes, but it depends on the average inclination, $\alpha$, of the
aligned spin axes. If $\alpha \leq 45^{\circ}$ then a well aligned
distribution produces a distinctive probability density that falls to
zero for $\lambda<90^{\circ}$. However, if $45< \alpha< 75^{\circ}$
then the probability density of $\lambda$ becomes quite flat, giving no
clear indication of the underlying distribution of spin axes.  The
simulations presented in section~3.4 also show that there are only
modest gains to be made by observing very large samples of stars
The exception would be a set of stars with a well-aligned distribution 
with a high mean inclination  ($\alpha \geq 75^{\circ}$). In this case
a relatively large sample ($\approx 100$ stars) is required to clearly
identify the high degree of alignment. 

The first application of this technique using data for 36 stars in the
Pleiades and Alpha Per is encouraging. For both clusters there is a clear
increase in probability density with cone angle, entirely consistent with 
random spin-axis orientation.  However, we cannot rule out partial
alignment of the spin axes, but place 90 per cent lower limits of about
$\lambda > 40^{\circ}$ for both clusters. Values of $\lambda$ that are
much less than $90^{\circ}$ could only be disguised in the data if 
the mean inclination were between $45^{\circ}$ and $75^{\circ}$ .  Our
results also show that the analysis method is relatively insensitive to
the exact levels of measurement uncertainty and small changes in the
assumed mean distance to the cluster. These results show that for the
method of analysis to be effective the average distance to the cluster
taken from independent measurements needs to be well defined with a
normalised uncertainty of a few per cent or better.


Should it turn out to be possible to assume that spin axes are always
randomly directed within a cluster, then this allows 
measurements of average stellar distances, radii and age spreads.
In this paper we have used this technique to derive new distance estimates
for the Pleiades and Alpha Per of $133\pm 7$\,pc and $182\pm 11$\,pc
respectively. These values are in good agreement with the main sequence
fitting distances to these clusters. The Pleiades result is marginally
higher than the Hipparcos parallax-based distance.  If larger samples
of stars could be measured, then unlike the test for random spin
alignment, there {\it are} significant gains (improving by a factor of
approximately $\sqrt{N}$) to be made in the precision of the mean $\sin
i$ value, which in turn determines the statistical precision of the
distance estimate. As there are far more $v \sin i$ measurements than
known periods in both the Pleiades and Alpha Per, it would be a
fruitful project to search for more rotation periods in both these
clusters. However, if spin-axis orientation cannot be assumed random, then
although the estimated distance remains unbiased when analysed under
the assumption of randomness, there are significant systematic errors
of up to $^{+18}_{-32}$ percent, which render this technique
ineffective as an independent means of estimating stellar radii
or cluster distance.

Well defined distances should soon become available for
many more open clusters following the launch of the GAIA satellite.
In addition periods are now being measured systematically in
nearby clusters (e.g. Aigrain et al. 2006). As these data become 
available it will be interesting to compare the distances derived
from satellite based parallax measurement with those estimated 
from the measured period and rotational velocity. If all results agree
within measurement errors then this would
support the general assumption \textit{that spin-axes are} randomly orientated in
space over the scale length of a cluster. This is because it
would become increasingly unlikely that the spins in all clusters were 
aligned, and had a mean inclination in the range $45^{\circ}$--$75^{\circ}$.
Conversely, if the spin axes in some clusters are well aligned there 
would be a reasonable chance ($\simeq 55$ per cent if we can assume 
that the mean inclination from cluster-to-cluster is a random variable) 
of observing a cluster with an average inclination of $\leq 45^{\circ}$ 
or $\geq 75^{\circ}$ in which case significant alignment could be 
identified using the method described in this paper.

\section*{Acknowledgements}
RJJ would like to thank the Science and Technology Facilities Council
for funding a postgraduate studentship.
 \nocite{Aigrain2007a}
 \nocite{An2007a}
 \nocite{Baxter2009a}
 \nocite{Jeffries2007a}
 \nocite{Jeffries2007b}
 \nocite{Hendry1993a}
 \nocite{Shu1987a}
 \nocite{Tamura1989a}
 \nocite{Vink2005a}
 \nocite{Menard2004a}
 \nocite{Cutri2003a}
 \nocite{ODell1994a}
 \nocite{Barnes1976a}
 \nocite{Stauffer2003a}
 \nocite{Kervella2004a}
 \nocite{Mermilliod1995a}
 \nocite{Rieke1985a}
 \nocite{Jackson2009a}
 \nocite{Soderblom2005a}

\bibliographystyle{mn2e} 
\bibliography{RJJbib}


\bsp 

\label{lastpage}

\end{document}